\documentclass{appolb}
\usepackage{graphicx}
\begin{document}
\title{$R$-dependence of inclusive jet suppression and groomed jet splittings in heavy-ion collisions with ALICE%
\thanks{Presented at the 29th International Conference on Ultrarelativistic Nucleus-Nucleus Collisions}%
}
\author{Hannah Bossi for the ALICE Collaboration
\address{Yale University}
\\[3mm]
}
\maketitle
\begin{abstract}
Jets in relativistic heavy-ion collisions (HICs) interact with the quark-gluon plasma (QGP), leading to effects such as a suppression of jet yields and a modification of internal jet structure that are used to measure the properties of the QGP. These proceedings show the inclusive jet nuclear modification factors in Pb--Pb collisions in various centrality classes at $\sqrt{s_{\rm NN}}$ = 5.02 TeV recorded with the ALICE detector for resolution parameters up to $R$ = 0.6 for momenta down to 40 GeV/$c$. The analysis utilizes machine learning techniques to correct the large background in HICs, extending the measurement of inclusive jets to lower jet $p_{T}$ and larger $R$ than previously achieved in HICs at the LHC. A new suite of measurements characterizing groomed jet splittings using both the Soft Drop and Dynamical Grooming algorithms in central and semi-central Pb--Pb collisions will also be presented. The groomed jet radius, $\theta_{\rm g}\equiv R_{\rm g}/R$, the groomed jet momentum fraction, $z_{\rm g}$, and the transverse momentum of the groomed splitting, $k_{\rm T, g}$ are also reported. All measurements are fully corrected through unfolding and compared to various theoretical calculations.
\end{abstract}
  
\section{Introduction}\label{sec:Introduction}
In HICs jets interact with the QGP, leading to modifications as compared to vacuum, which is generally referred to as \textit{jet quenching}. Experimental expectations of jet quenching can be grouped into three categories. One expectation of jet quenching is that parton energy loss will lead to a suppression of jet yields in HICs compared to the yield if no QGP were present, an effect referred to as jet suppression. Jet suppression is commonly measured via the jet nuclear modification factor ($R_{\rm AA}$) given by $   R_{\rm AA} = \frac{\frac{1}{\langle T_{\rm AA} \rangle}\frac{1}{N_{\rm events, PbPb}} \frac{d^{2}N^{\rm AA}_{\rm jet}}{dp_{\rm T}d\eta}}{\frac{d^{2}\sigma_{\rm pp}}{dp_{\rm T}d\eta}}$.
The second category of jet quenching signatures is the modification of jet substructure. This modification can be observed experimentally via effects such as momentum broadening or the addition of soft particles due to a medium-induced wake. Additionally, the third category of jet quenching expectations is a deflection of the jet centroid due to multiple soft scatterings or scatterings with QGP quasi-particles. Each of these signatures may modify jets differently depending on their partonic structures, flavors, $p_{\rm T}$, path lengths through the plasma and other effects. For example, gluon jets are expected to lose more energy than quark jets due to their respective color charges. The same jets can also lose energy differently due to fluctuations in jet-medium interactions. Therefore, a critical component in jet quenching measurements is to effectively characterize the jet population in order to disentangle these competing energy loss mechanisms. Another important component of experimental jet measurements is to mitigate the large background in HICs. In these proceedings, three recent ALICE measurements will be discussed, each of which is designed to probe a different expectation of jet quenching. 

\section{Jet Splittings}\label{sec:Jetsplittings}

Measurements of jet splittings utilize the declustering history in order to probe partonic splittings within the jet and their subsequent modification. The asymmetry present in a jet splitting is characterized by the shared momentum fraction, $z =  \frac{p_{\rm T, 2}}{p_{\rm T, jet}}$, between the subjets. Here $p_{\rm T, 2}$, $p_{\rm T, jet}$ refer to the $p_{\rm T}$ of the subleading subjet and the jet, respectively. The hardness of a splitting is characterized by the relative $p_{\rm T}$ of the subjets, defined as $k_{\rm T} = p_{\rm T, 2}\mathrm{sin}(\Delta R)$. The width of a splitting is given by the opening angle between subjets, $\theta = \frac{\Delta R}{R}$. 

To isolate the hard partonic splittings of the jet, \textit{grooming methods} are commonly applied. In these proceedings, results with Soft Drop (SD) \cite{Larkoski:2014wba} and dynamical grooming (DyG) \cite{PhysRevD.101.034004} will be discussed. In the SD grooming procedure, hard splittings are selected by a $z$ cutoff given by $z > z_{\rm cut}(\frac{\Delta R}{R})^{\beta}$, where results in these proceedings all use $\beta$ = 0. These results use $z_{\rm cut} = 0.2$, which is shown to help mitigate the heavy-ion background \cite{PhysRevC.102.044913}. Unlike SD, dynamical grooming seeks to find the hardest branch amongst an iterative set of splittings. Here the grooming cutoff in $z$ is not a fixed number (as in SD grooming), but is generated on a jet-by-jet basis as specified in Eq.~\ref{eq:dynamicalgrooming}. Different values for $a$ correspond to different hardness metrics. 

\begin{equation}\label{eq:dynamicalgrooming}
    \kappa^{(a)} = \frac{1}{p_{\rm T}} \mathrm{max}_{i \in \mathrm{C/A}}z_{\rm i}(1-z_{\rm i})p_{\rm T, i}(\frac{\Delta R_{\rm i}}{R_{\rm i}})^{a}
\end{equation}

In HICs, jet splittings are used to probe the modification of jet substructure as well as the resolution length of the medium and its spacetime structure. When the resolution length of the medium is zero, this is referred to as the fully decoherent limit, where the medium can resolve all splittings. In this case, each subjet loses energy independently, resulting in more overall energy loss. When the resolution length of the medium is infinite, the medium is unable to resolve the jet splitting and the jet loses energy as a single object. This case, referred to as the fully coherent limit, results in less overall energy loss as compared to the fully decoherent limit. 

ALICE has performed a measurement of the groomed jet radius ($\theta_{\rm g} = \frac{R_{\rm g}}{R}$) at 5.02 TeV in pp and Pb--Pb, which is shown in the left panel of Fig.~\ref{fig:hardestKT}. The ratio between Pb--Pb and pp is additionally shown and indicates a suppression of wide-angle splittings in Pb--Pb relative to pp collisions, or a narrowing effect. When compared to jet quenching models \cite{Casalderrey_Solana_2020, Caucal_2019, Ringer:2019rfk, Putschke:2019yrg} this is favored by models including decoherence\cite{Casalderrey_Solana_2020}. These results also are consistent with models including coherence with a high quark fraction\cite{Ringer:2019rfk}. For more details see Refs. \cite{ALICE:2022eoc,ALargeIonColliderExperiment:2021mqf}.

\section{Hardest $k_{\rm T}$}\label{sec:hardestKT}
Jet splitting techniques can also be useful in order to search for point-like or Moliere scatterings in the QGP\cite{Barata:2021wuf,DEramo:2018eoy}. Moliere scatterings would result in an excess of large $k_{\rm T, g}$ (groomed $k_{\rm T}$) splittings in Pb--Pb collisions as compared to pp collisions. This observable is additionally sensitive to the modification of the jet's internal structure. A variety of SD and dynamical grooming methods can be used to identify the hardest $k_{\rm T, g}$ splitting. 

\begin{figure}[htb!]
    \centering
     \includegraphics[width = 0.3\textwidth]{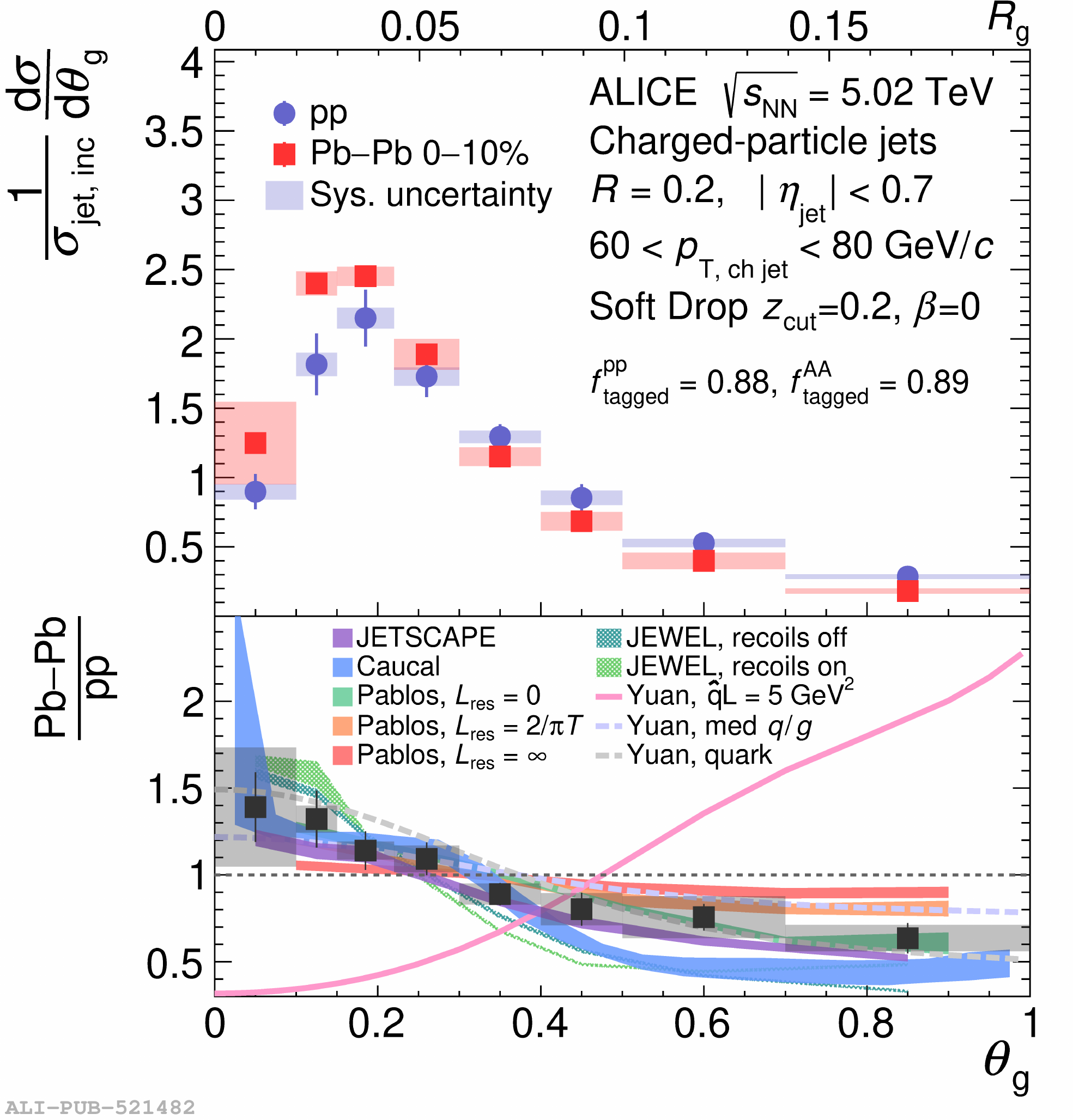}
    \includegraphics[width = 0.3\textwidth]{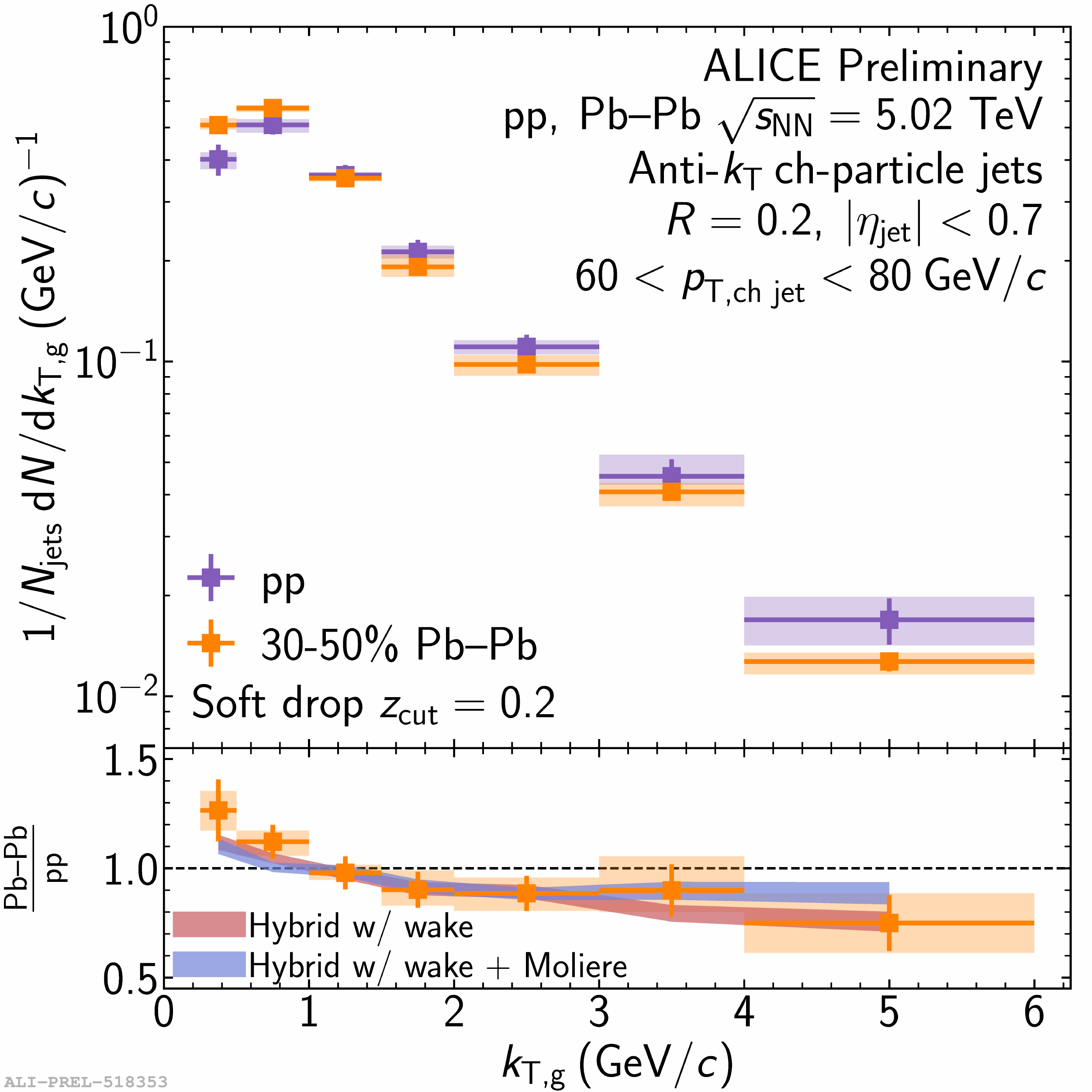}
    \includegraphics[width = 0.3\textwidth]{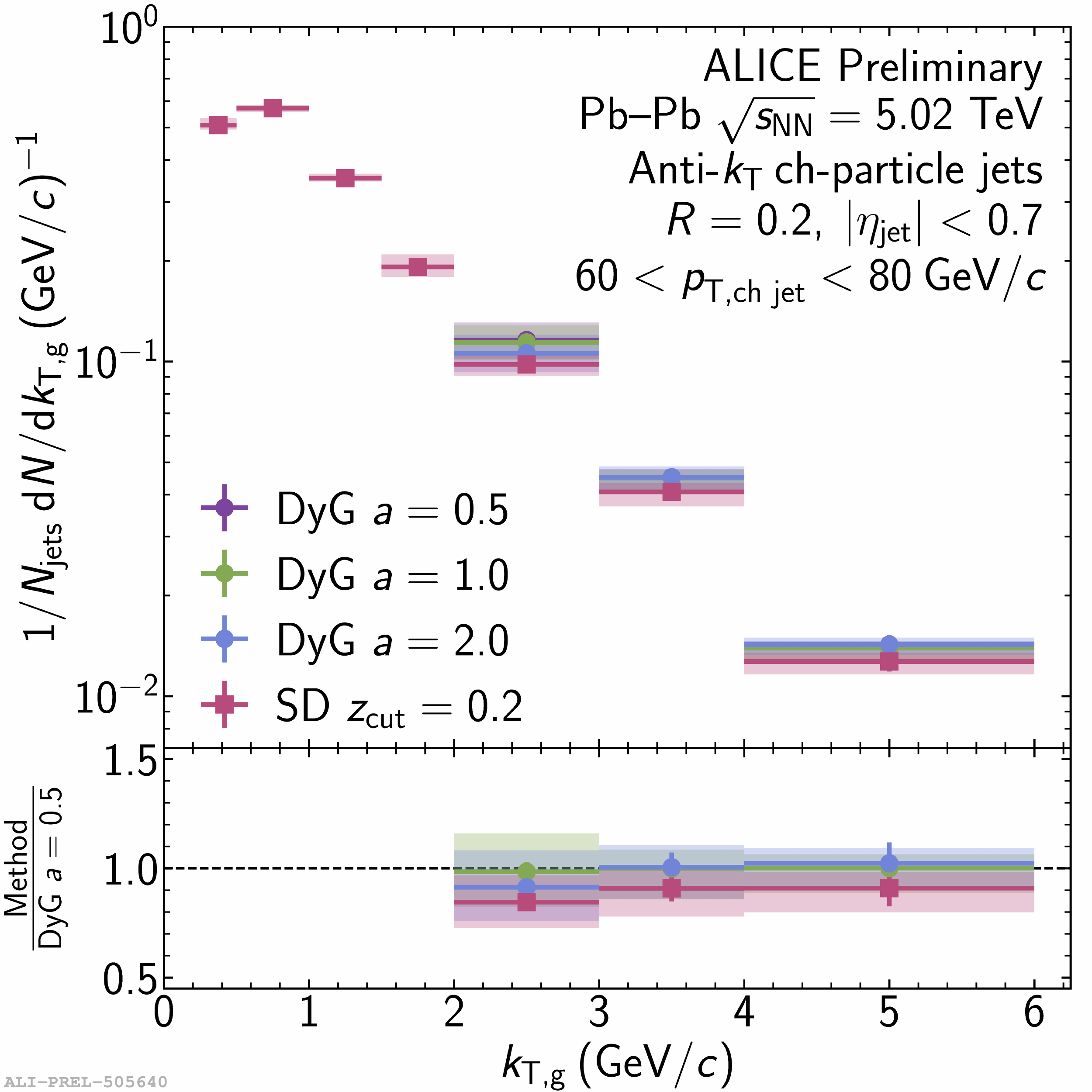}
    \caption{Unfolded $\theta_{\rm g}$ distributions in pp collisions as compared to central (0-10\%) Pb--Pb collisions (Left). The ratio is then compared to model predictions.(Middle) Comparison of the SD hardest $k_{\rm T, g}$ jet splitting found for $R = 0.2$ jets in pp and semi-central Pb--Pb collisions. The ratio is compared with hybrid model predictions with and without Moliere scattering. (Right) Grooming methods are compared for $R$ = 0.2 jets in semi-central Pb--Pb collisions.}
    \label{fig:hardestKT}
\end{figure}

The distributions of the hardest $k_{\rm T, g}$ in pp and 30-50\% Pb--Pb collisions are shown in Fig.~\ref{fig:hardestKT}. The ratio of these distributions does not show the expected enhancement of large $k_{\rm T, g}$ splittings and instead is consistent with no modification at high $k_{\rm T, g}$. However, note that the model predictions \cite{DEramo:2018eoy}, both with and without Moliere effects are consistent with the data in this region. At low $k_{\rm T, g}$, there is a hint of a modification that is both consistent with the narrowing effect observed in $R_{\rm g}$ and described by the model predictions. Additionally, the results with the various SD and dynamical grooming techniques are consistent within uncertainties, as shown in Fig.~\ref{fig:hardestKT}.  

\section{$R$-dependence of the $R_{\rm AA}$}\label{sec:Raa}
The $R$-dependence of the $R_{\rm AA}$ is very useful when discriminating between different competing energy loss mechanisms, each of which may cause the $R_{\rm AA}$ to increase or decrease with $R$. For example, the recovery of wide-angle radiation with increasing $R$ will cause the $R_{\rm AA}$ to increase with $R$. The medium response can also add energy to the jet cone, increasing the $R_{\rm AA}$ with $R$. However, large $R$ jets may have more effective energy-loss sources, resulting in a decrease in the $R_{\rm AA}$ with $R$. The quark-to-gluon ratio at a fixed $p_{\rm T}$ will also increase for larger values of $R$, which would cause the $R_{\rm AA}$ to decrease with $R$. Each of these mechanisms can be present in different amounts simultaneously for a given population of jets, making it important to measure over wide ranges in jet $p_{\rm T}$ and $R$. However, inclusive jet measurements at large $R$ and lower $p_{\rm T}$ are made difficult by the large fluctuating underlying event.

\begin{figure}[htb!]
    \centering
    \includegraphics[width = 0.3\textwidth]{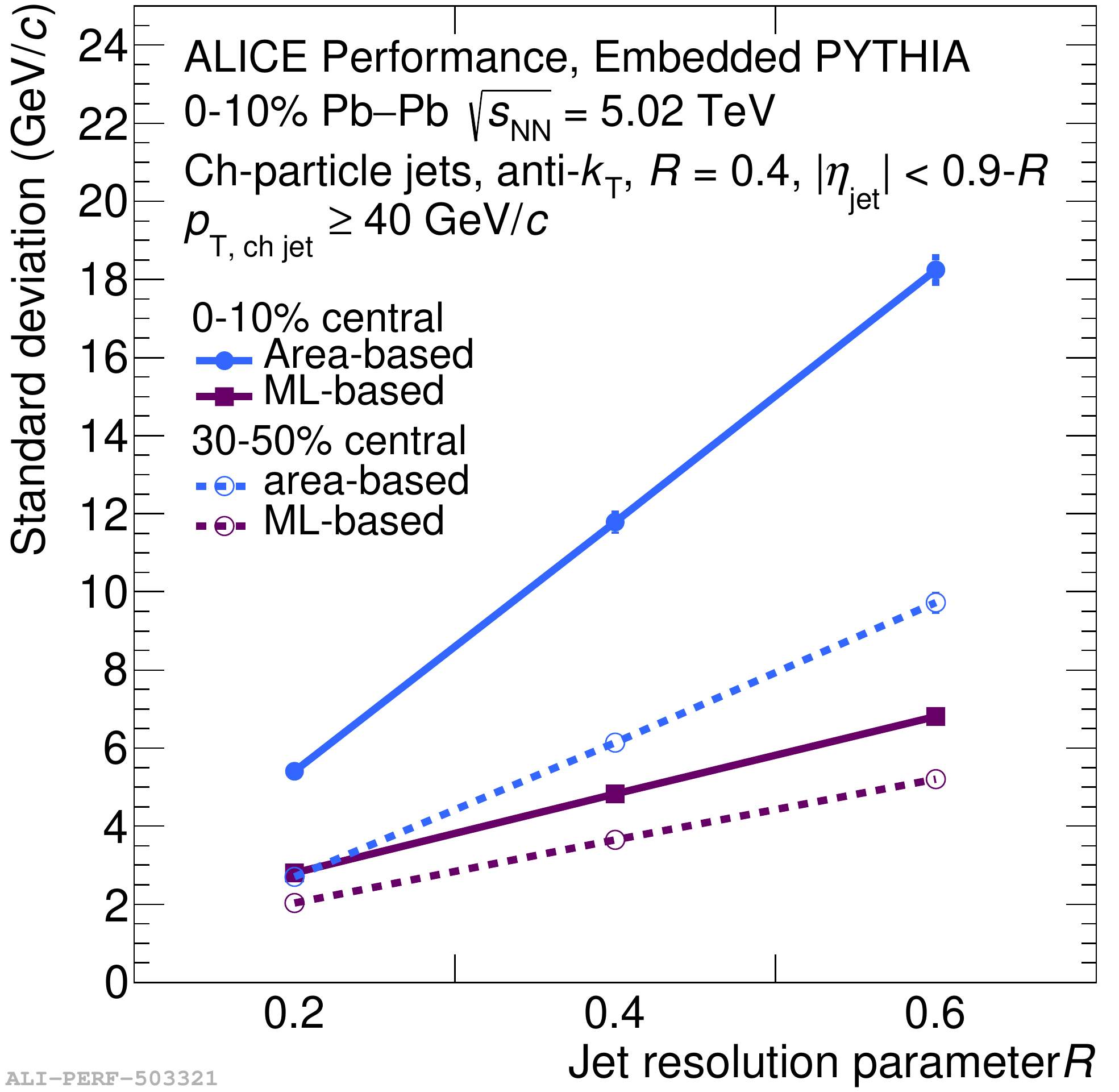}
    \includegraphics[width = 0.3\textwidth]{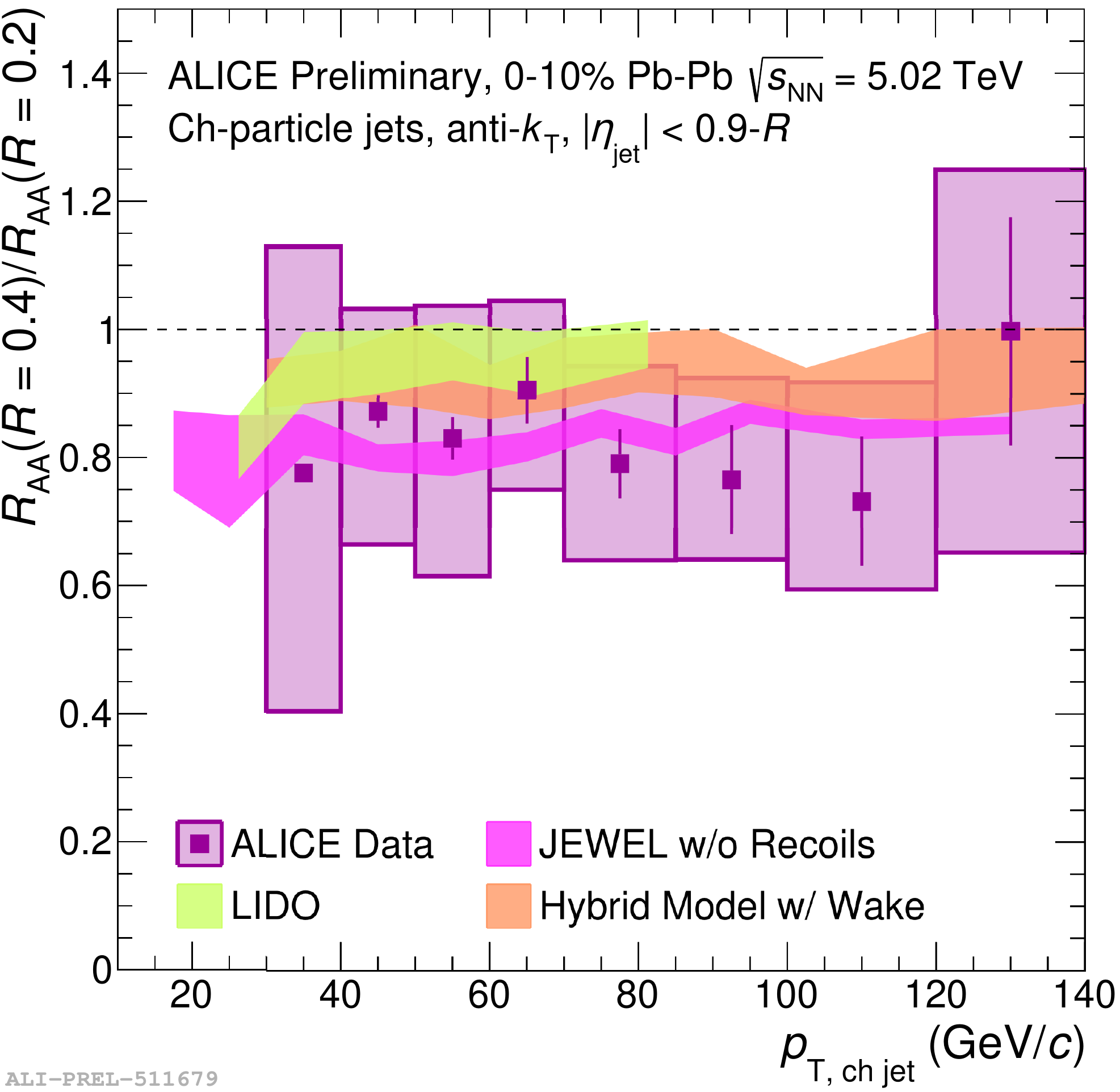}
    \includegraphics[width = 0.3\textwidth]{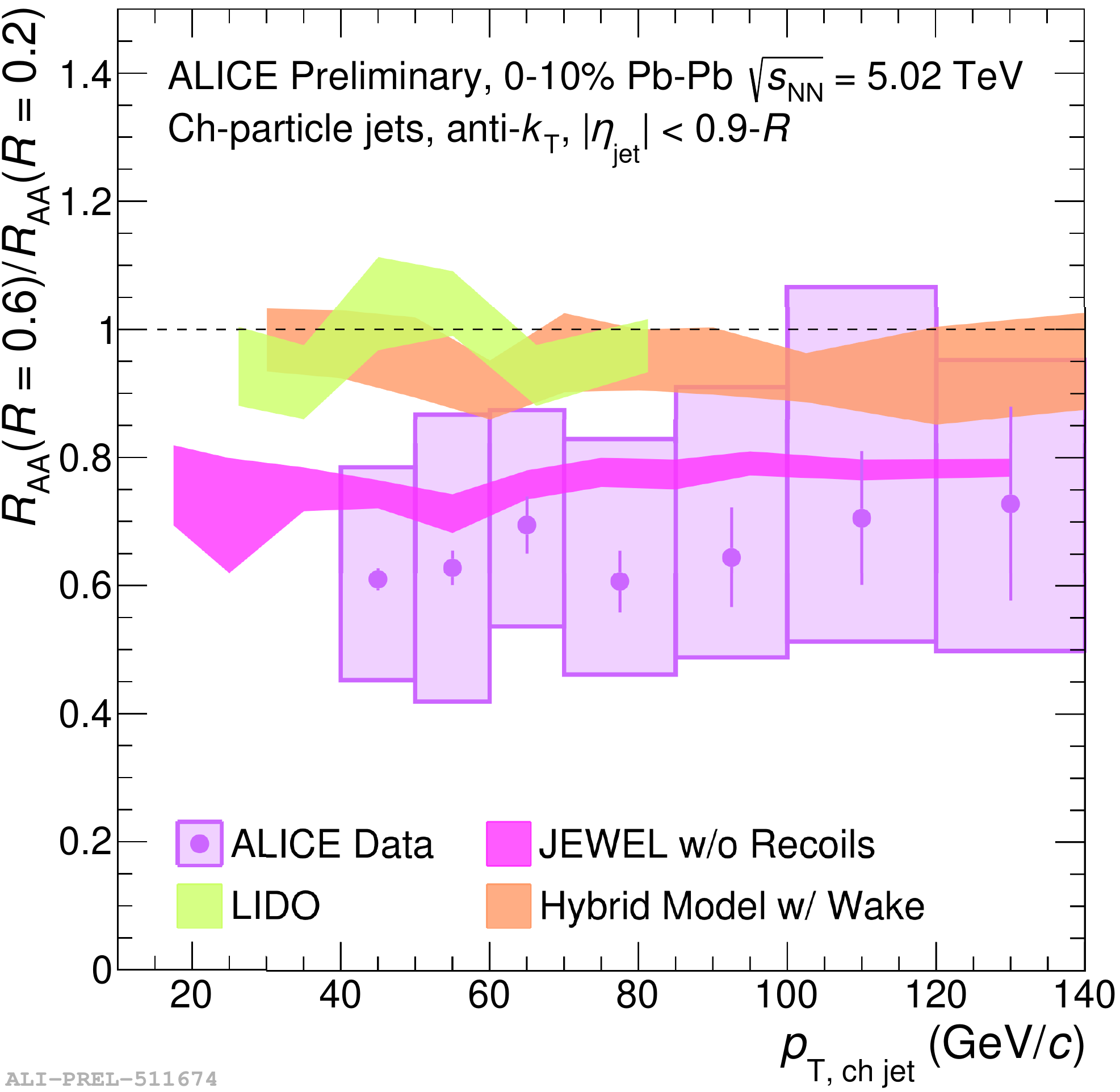}
    \caption{Standard deviation of the $\delta p_{\rm T} = p_{\rm T, rec} - p_{\rm T, true}$ as a function of $R$ (left). Double ratio of jet $R_{\rm AA}$ using $R$ = 0.4 (middle) and $R$ = 0.6 (right) as the numerator and $R=0.2$ as the denominator compared to model predictions. }
    \label{fig:raaratios}
\end{figure}

The results presented in these proceedings utilize a novel machine learning (ML) based background estimator in order to correct for the background in HICs \cite{Haake:2018hqn}. Here, the ML estimator learns a mapping between the measured and corrected jet from selected jet properties, including properties of the jet constituents. The inclusion of constituent information causes a dependence on the fragmentation of the jet used in training. As the ML is trained on PYTHIA, and the fragmentation of jets in HICs is known to be modified \cite{ATLAS:2018bvp,CMS:2014jjt}, this dependence must be accounted for. To do this, three toy modifications to the fragmentation used in training are applied. These modifications mimic the effects of in-cone radiation, out-of-cone radiation, and the medium response recreated with data-motivated parameters \cite{CMS:2018zze}. From these variations a systematic uncertainty is derived for the final result. The ML-based background estimator exhibits significant improvement over the standard ALICE background subtraction method, as evidenced by a reduction in the standard deviation of the $\delta p_{\rm T}$ distribution as shown in Fig.~ \ref{fig:raaratios}. These improvements are largest in central collisions and at large $R$ which, after applying an unfolding procedure for residual background and detector effects, significantly extend the kinematic reach in these regimes.

The ratios of the $R_{\rm AA}$ at different $R$ values as compared to $R$ = 0.2 ($R_{\rm AA} (R)/ R_{\rm AA} (R = 0.2)$) probe the $R$-dependence of the $R_{\rm AA}$, by revealing if the $R_{\rm AA}$ increases (ratio above 1) or decreases (ratio below 1) with increasing $R$. The $R_{\rm AA}$ ratios for 0-10\% collisions are shown in Figure \ref{fig:raaratios} as compared to jet quenching models \cite{Ke:2020clc,KunnawalkamElayavalli:2017hxo, Pablos:2019ngg}, where the middle panel is the $R_{\rm AA} (R = 0.4)/ R_{\rm AA} (R = 0.2)$ and the right panel is the $R_{\rm AA} (R = 0.6)/ R_{\rm AA} (R = 0.2)$ ratio. The $R_{\rm AA} (R = 0.4)/ R_{\rm AA} (R = 0.2)$ and the $R_{\rm AA} (R = 0.6)/ R_{\rm AA} (R = 0.2)$ ratio are below unity, indicating an $R$ dependence. For the $R_{\rm AA} (R = 0.4)/ R_{\rm AA} (R = 0.2)$ the deviation from unity is not significant and in $R_{\rm AA} (R = 0.6)/ R_{\rm AA} (R = 0.2)$ ratio the deviation from unity is marginally significant. Therefore, effects like wider jets losing more energy or changes in the q/g fraction have a dominant influence compared to other effects such as the recovery of energy lost or the medium response. These trends are roughly consistent with the included predictions with some tension in the large $R$ ratio. These results are additionally consistent with the conclusion from the $R_{\rm g}$ and $k_{\rm T, g}$ measurements of a narrowing of the inclusive jet distribution in Pb--Pb as compared to vacuum.

\vspace{-5mm}

\section{Conclusions}\label{sec:Conclusions}
These proceedings outline three different measurements, each probing a different signature of jet quenching. Jet splittings are a useful technique in order to isolate hard splittings within a jet and to probe the modification of jet substructure. Measurements of the groomed jet radius indicate a suppression of wide-angle splittings in Pb--Pb collisions as compared to pp collisions. ALICE has also investigated jet suppression via the $R$-dependence of the $R_{\rm AA}$. A decrease in the $R_{\rm AA}$ is observed, where larger $R$ jets are more suppressed. Jet splitting techniques are additionally used to measure the hardest $k_{\rm T, g}$ splittings, where a modification in this distribution is seen from pp to Pb--Pb collisions. Overall, these measurements are consistent with a narrowing of the inclusive jet distribution in HICs at a fixed jet $p_{\rm T}$.   

{

\renewcommand{\refname}{}
\vspace{-40pt}
\bibliography{refs} }

\begin{thebibliography}{10}

\bibitem{ALICE:2022eoc}
{Physics Preliminary Summary: Measurements of the groomed jet radius and
  momentum splitting fraction with the soft drop and dynamical grooming
  algorithms in pp collisions at $\sqrt{s}=5.02$ TeV}.
\newblock 2022.

\bibitem{ATLAS:2018bvp}
M.~Aaboud et~al.
\newblock {Measurement of jet fragmentation in Pb+Pb and $pp$ collisions at
  $\sqrt{s_{NN}} = 5.02$ TeV with the ATLAS detector}.
\newblock {\em Phys. Rev. C}, 98(2):024908, 2018.

\bibitem{ALargeIonColliderExperiment:2021mqf}
S.~Acharya et~al.
\newblock {Measurement of the groomed jet radius and momentum splitting
  fraction in pp and Pb$-$Pb collisions at $\sqrt{s_{NN}} = 5.02$ TeV}.
\newblock {\em Phys. Rev. Lett.}, 128(10):102001, 2022.

\bibitem{Barata:2021wuf}
J.~a. Barata, Y.~Mehtar-Tani, A.~Soto-Ontoso, and K.~Tywoniuk.
\newblock {Medium-induced radiative kernel with the Improved Opacity
  Expansion}.
\newblock {\em JHEP}, 09:153, 2021.

\bibitem{Casalderrey_Solana_2020}
J.~Casalderrey-Solana, G.~Milhano, D.~Pablos, and K.~Rajagopal.
\newblock Modification of jet substructure in heavy ion collisions as a probe
  of the resolution length of quark-gluon plasma.
\newblock {\em Journal of High Energy Physics}, 2020(1), jan 2020.

\bibitem{Caucal_2019}
P.~Caucal, E.~Iancu, and G.~Soyez.
\newblock Deciphering the zg distribution in ultrarelativistic heavy ion
  collisions.
\newblock {\em Journal of High Energy Physics}, 2019(10), oct 2019.

\bibitem{CMS:2014jjt}
S.~Chatrchyan et~al.
\newblock {Measurement of Jet Fragmentation in PbPb and pp Collisions at
  $\sqrt{s_{NN}}= 2.76$ TeV}.
\newblock {\em Phys. Rev. C}, 90(2):024908, 2014.

\bibitem{DEramo:2018eoy}
F.~D'Eramo, K.~Rajagopal, and Y.~Yin.
\newblock {Moli\`ere scattering in quark-gluon plasma: finding point-like
  scatterers in a liquid}.
\newblock {\em JHEP}, 01:172, 2019.

\bibitem{Haake:2018hqn}
R.~Haake and C.~Loizides.
\newblock {Machine Learning based jet momentum reconstruction in heavy-ion
  collisions}.
\newblock {\em Phys. Rev. C}, 99(6):064904, 2019.

\bibitem{Ke:2020clc}
W.~Ke and X.-N. Wang.
\newblock {QGP modification to single inclusive jets in a calibrated transport
  model}.
\newblock {\em JHEP}, 05:041, 2021.

\bibitem{KunnawalkamElayavalli:2017hxo}
R.~Kunnawalkam~Elayavalli and K.~C. Zapp.
\newblock {Medium response in JEWEL and its impact on jet shape observables in
  heavy ion collisions}.
\newblock {\em JHEP}, 07:141, 2017.

\bibitem{Larkoski:2014wba}
A.~J. Larkoski, S.~Marzani, G.~Soyez, and J.~Thaler.
\newblock {Soft Drop}.
\newblock {\em JHEP}, 05:146, 2014.

\bibitem{PhysRevD.101.034004}
Y.~Mehtar-Tani, A.~Soto-Ontoso, and K.~Tywoniuk.
\newblock Dynamical grooming of qcd jets.
\newblock {\em Phys. Rev. D}, 101:034004, Feb 2020.

\bibitem{PhysRevC.102.044913}
J.~Mulligan and M.~P\l{}osko\ifmmode~\acute{n}\else \'{n}\fi{}.
\newblock Identifying groomed jet splittings in heavy-ion collisions.
\newblock {\em Phys. Rev. C}, 102:044913, Oct 2020.

\bibitem{Pablos:2019ngg}
D.~Pablos.
\newblock {Jet Suppression From a Small to Intermediate to Large Radius}.
\newblock {\em Phys. Rev. Lett.}, 124(5):052301, 2020.

\bibitem{Putschke:2019yrg}
J.~H. Putschke et~al.
\newblock {The JETSCAPE framework}.
\newblock 3 2019.

\bibitem{Ringer:2019rfk}
F.~Ringer, B.-W. Xiao, and F.~Yuan.
\newblock {Can we observe jet $P_T$-broadening in heavy-ion collisions at the
  LHC?}
\newblock {\em Phys. Lett. B}, 808:135634, 2020.

\bibitem{CMS:2018zze}
A.~M. Sirunyan et~al.
\newblock {Jet properties in PbPb and pp collisions at $
  \sqrt{s_{\mathrm{N}\;\mathrm{N}}}=5.02 $ TeV}.
\newblock {\em JHEP}, 05:006, 2018.

\end{thebibliography}

\end{document}